\documentstyle[preprint,aps]{revtex}
\def\Z{{\cal Z}}
\def\rv{{\bf r}}
\def\lb{\langle}
\def\rb{\rangle}

\begin{document}

\title{Monte Carlo study of the Widom-Rowlinson fluid using cluster
methods}

\author{Gregory Johnson and Harvey Gould}
\address{Department of Physics, Clark University,
Worcester, MA 01610-1477}

\author{J. Machta}
\address{Department of Physics and
Astronomy, University of Massachusetts, Amherst, MA 01003-3720}

\author{L. K. Chayes}
\address{Department of Mathematics, University of California, Los
Angeles, CA 90095-1555}

\maketitle
\begin{abstract}
The Widom-Rowlinson model of a fluid mixture is studied using a new
cluster algorithm that is a generalization of the invaded cluster
algorithm previously applied to Potts models. Our estimate of the
critical exponents for the two-component fluid are consistent with
the Ising universality class in two and three dimensions. We also
present results for the three-component fluid.
\end{abstract}

Some years ago Widom and Rowlinson~\cite{WiRo} introduced a simple
but non-trivial continuum model that exhibits a phase
transition~\cite{R}. The two-component formulation of
this model consists of ``black'' and ``white'' particles; particles
of the same type do not interact, but particles of differing type
experience a hard-core repulsion at separations less than or
equal to $\sigma$~\cite{note}. In this letter we present a new Monte Carlo method
for simulating the Widom-Rowlinson (WR) model and apply the method
to study the properties of the transition point in two and three
dimensions.

There have been relatively few Monte Carlo studies of the WR critical
point because of the combined difficulties of treating hard-core
systems and critical
slowing down using standard Monte Carlo techniques.
The algorithm presented here overcomes these
difficulties by using a generalization of the cluster
approach introduced by Swendsen and Wang~\cite{SwWa}. The
algorithm is a variant of the invaded cluster
(IC) method~\cite{MaCh95,MaCh96} adapted to continuum problems. We
find that the algorithm has almost no critical slowing down and that
we can obtain accurate values of the critical density and the
exponent ratios
$\beta/\nu$ and $\gamma/\nu$ with relatively modest computational
effort.

The 2-component WR model is expected to be
in the Ising universality class. Our results for
$\beta/\nu$ and $\gamma/\nu$ are consistent with this assumption and
are in good agreement with the best known values for the Ising
model. Our value for the critical density of the three-dimensional
($d=3$) WR model agrees with recent results obtained by
Shew and Yethiraj~\cite{ShYe}.

We also consider a WR model in which
there are $q$ components, any two of which interact via a hard-core
repulsion. Despite the apparent similarity to the
$q$-state Potts model, the phase structure may have additional
features~\cite{RL,CKS}.  Our
algorithm easily extends to these
$q$-component WR models, and we present results for the
3-component model in $d=2,3$.

\paragraph*{Graphical Representations of the WR Model.}
A configuration of the WR fluid
consists of two sets of points, $S$ and $T$, corresponding to the
positions of the black and white particles
respectively. In the grand canonical ensemble, the probability
density for finding the configuration $(S,T)$ is
\begin{equation}
\label{eq:WR}
P(S,T)=\frac{1}{\Z}\frac{z_1^{N_1}}{N_1!}
\frac{z_2^{N_2}}{N_2!} \,
\Gamma(S,T).
\end{equation}
$\Z$ is the grand partition
function, $z_1$ ($z_2$) is the fugacity of the black (white)
particles, and $N_1$ ($N_2$) is the number of black (white)
particles. The object $\Gamma$, which expresses the hard-core
interaction between particles of different types, vanishes if
any point in
$S$ is within a distance $\sigma$ of any point in $T$ and is one
otherwise.  Symmetry considerations insure that the critical point is
along the line $z_1=z_2$; hereafter we restrict attention to
the case, $z=z_1=z_2$.

To motivate and justify our cluster algorithm, we consider
a different representation of the WR model. In the ``gray''
representation~\cite{CCK,GML}
one considers the particles without reference to color, but
with configurations weighted according to the number of
allowed colorings. Let $W$ be a list of $N$ points, $W
\equiv (\rv_1, \dots \rv_N)$.
Clusters of particles can be defined by the condition that every
particle in a cluster is within a distance $\sigma$ of some
other particle in the cluster. Particles in a cluster must all be
the same color, so if there are  $C(W)$ distinct clusters (including
single particles), there are $2^{C(W)}$ allowed colorings. Starting
with Eq.~(\ref{eq:WR}) and working through the combinatorics, we
find that the probability density for $W$ is given by
\begin{equation}
\label{eq:fk} p(W)= \frac{1}{\Z} \frac{z^N}{N!}
2^{C(W)}.
\end{equation} These densities describe the {\em gray
measures}. The appropriate gray measure for the $q$-component model
is defined by the analog of Eq.~(\ref{eq:fk}) with 2 replaced by
$q$. To return to the distribution in Eq.~(\ref{eq:WR}) starting
from the gray representation, we select one of the
$2^{C(W)}$ (or $q^{C(W)}$) allowed colorings with equal probability.
It turns out that the gray measures are a special case of the models
studied in~\cite{CoKl}. The connection with the WR models was
discussed in~\cite{GS} and made precise in~\cite{CCK,GML}.

\paragraph*{Cluster Algorithms.}
With the gray representation in mind, we consider the following
cluster algorithm. Starting from a configuration $W$ of gray
particles, clusters are identified. Each cluster is independently
labeled black or white with probability
1/2 and all the white particles are removed. In the next step, white
particles are replaced via a Poisson process at fugacity
$z$ in the free volume permitted by the black particles. In the
final step, color identities are erased and we obtain the next gray
configuration. The generalization to the
$q$-component WR model is that the fraction of clusters
deleted is $1/q$. This algorithm is the
generalization of the Swendsen-Wang method to the WR model.  The
algorithm is described in more detail and detailed balance is proved
in~\cite{CM} and independently in~\cite{HLM}.

Because we are interested in efficiently sampling the
transition point of WR models, we will forsake a
Swendsen-Wang approach in favor of an IC algorithm. The steps
involving the coloring and discarding of clusters are essentially the
same, but rather than repopulating the free volume by a fixed
fugacity process, particles are sequentially added with a uniform
distribution throughout the free volume until a {\em stopping
condition} is fulfilled. For example, one could add particles until
a fixed particle number is reached. If a stopping rule is chosen
that enforces a condition which is characteristic of criticality,
then a critical state of the system is sampled automatically. In a
finite volume the IC method samples an ensemble that differs
from the canonical ensemble, but it presumably converges to the
correct infinite volume distribution for all local observables. The
validity of the IC method and its relation to the
Swendsen-Wang method is discussed briefly below and in detail in
Ref.~\cite{MaCh96}.

The signature of the phase transition in the WR model is
percolation of a gray cluster~\cite{CCK,GML}. Thus an
appropriate stopping rule for the IC algorithm is the spanning of a
gray cluster -- particles are added at random in the allowed volume
until a cluster spans the system. (In our case, we use
periodic boundary conditions and spanning is said to occur when a
cluster wraps the torus.) The other modification is that the
spanning cluster is erased on each deletion move, thereby ensuring
that a new spanning cluster can form in the repopulation move.

If $N_{\rm tot}$ denotes the total number of points that are needed
to satisfy the stopping condition, then
$z=N_{\rm tot}/V$ is an estimator for the critical fugacity
$z_c$. In the limit $V \rightarrow \infty$, it is reasonable to
assume that the distribution for $z$ becomes sharp. If this
assumption is valid, each move is identical to a move of the
Swendsen-Wang-type algorithm at the peak value of $z$. It follows
that this peak value is $z_c$ -- no other value of the fugacity
would exhibit a critical cluster. Hence, if the distribution for $z$
is very narrow in a finite volume, the IC algorithm is essentially
the Swendsen-Wang-type algorithm with small fluctuations in the
fugacity.

\paragraph*{Results.}
We now present our simulation results using
the IC method. We collected statistics for the following quantities:
the average number of particles in the spanning
cluster, $M$; the normalized autocorrelation function $\Gamma_M(t)$
of the number of particles in the spanning cluster as a function of
``time'' $t$ as measured in Monte Carlo steps; the compressibility
$\chi$ defined by
\begin{equation}
\label{chi}
\chi = {1 \over V} \! \sum_i s_i^2 ,
\end{equation}
where $s_i$ is the mass of the $i$th cluster and the spanning cluster
is included in this sum; the estimator for the critical fugacity
$z=\lb N_{\rm tot}\rb/V$; the fluctuations in $z$, $\sigma_z^2 = (\lb
N_{\rm tot}^2\rb-\lb N_{\rm tot}\rb^2)/V^2$; and the average number
of gray particles per unit area (volume) $\rho$ which is an
estimator of the critical density.  System size is measured in
units of the particle diameter, $\sigma$.  The results for these
quantities are presented in Tables I--IV along with our estimate of
the statistical errors.

The critical exponents that depend on the magnetic exponent $y_h$
can be obtained from the fractal dimension of the spanning cluster
via $M \sim L^{D}$ or from the compressibility via $\chi \sim
L^{\gamma/\nu}$.  The various
critical exponents are related by $D=y_h$,
$\gamma/\nu= 2 y_h-d$, and
$\beta/\nu=d-y_h$. The dynamical properties of the algorithm can be
measured by the integrated autocorrelation time defined by
$\tau_M = {1 \over 2} + \sum_{t=1}^\infty \Gamma_M(t)$.
The integrated autocorrelation time is roughly the number of Monte
Carlo steps between statistically independent samples and enters into
the error estimate for $M$. In practice, it is necessary to cut off
the upper limit of the sum defining $\tau_M$ when $\Gamma_M$ becomes
comparable to its error. The increase in $\tau_M$ defines a dynamic
exponent $z_M$ via $\tau_M \sim L^{z_M}$.

Our results for the 2-component,
$d=2$ WR fluid are summarized in Table~\ref{d2q2}. From the
log-log plot of
$M$ versus $L$ shown in Fig.~\ref{fig:MLq2d2}, we find that $y_h=d -
\beta/\nu=1.873 \pm 0.002$, and hence $\beta/\nu \approx 0.127$, a value
consistent with the exact Ising result of $\beta/\nu=1/8$.
Similarly, a log-log plot of
$\chi$ versus $L$ yields $\gamma/\nu=1.743 \pm 0.003$, consistent
with the exact Ising value, $\gamma/\nu = 7/4$.  These
results support the hypothesis that the 2-component $d=2$ WR fluid is
in the Ising universality class. The errors quoted here are
associated with the least squares fitting procedure and are two
standard deviations. We also estimated the statistical errors in
$y_h$ and $\gamma/\nu$ by generating synthetic data sets consistent
with the estimated errors in the measured values of
$M$ and $\chi$ and found similar results.

From the data of Table \ref{d2q2} we have estimated the infinite
volume critical values of the density, $\rho_c$ and fugacity,
$z_c$. A linear fit for $\rho(L)$ versus $1/L$ yields
$\rho_c=1.5652$. A three parameter fit of the form
\begin{equation}
\label{eq:fss22}
\rho(L) = \rho_c - A/L^{x} ,
\end{equation}
yields $\rho_c=1.5662$ with $x=0.96$. Because the biggest source of
error is the uncertainty in the fitting form rather than the
statistical errors in the raw data, we estimate the error in
$\rho_c$ as several times the difference between these two fits.
Hence, we conclude that $\rho_c=1.566 \pm 0.003$.

Within the statistical error the fugacity $z$ is unchanged for the
three largest system sizes. We take these values and several times
the statistical error to estimate the critical fugacity,
$z_c= 1.726 \pm 0.002$. To our knowledge, there are no independent
estimates of $\rho_c$ and $z_c$ for the 2-component,
$d=2$ WR fluid.

The autocorrelation function, $\Gamma_M(t)$ decreases
rapidly and oscillates about zero after
$t\approx 10$. Our results for $\tau_M$ for various system sizes are
summarized in Table~\ref{d2q2}. The slow increase of $\tau_M$ with
$L$ indicates that the dynamic critical exponent $z_M$ is small or
zero ($\tau_M \sim \ln L$). Because of its small value and our
limited data, we cannot make a
more precise statement.   Fitting $\Gamma_M(t)$ to a single
exponential leads to decorrelation times similar to the integrated
autocorrelation time.

Our results for the
2-component, $d=3$ WR fluid are summarized in
Table~\ref{d3q2}. Power law fits of
$M$ and $\chi$ versus $L$ yield $y_h = 3-\beta/\nu=2.479 \pm 0.001$
and $\gamma/\nu=1.961 \pm 0.003$, respectively. These values are
consistent with each other and with the recent estimate
of $y_h = 2.4815(15)$ obtained in Ref.~\cite{BlLuHe}.
The results confirm the expectation that the 2-component,
$d=3$ WR model is in the
$d=3$ Ising universality class.

In Fig.~\ref{fig:rLq2d3} we show $\rho$ versus $1/L$.
A linear fit yields $\rho_c=0.7484$. A three parameter fit of the
form given in Eq.~(\ref{eq:fss22})
yields $\rho_c=0.7478$ with $x=1.16$. We estimate the error
in $\rho_c$ as several times the difference between these two fits
and conclude that $\rho_c=0.748 \pm 0.002$. This value of
$\rho_c$ is in agreement with and improves upon the recent result in
Ref.~\cite{ShYe},
$\rho_c=0.762 \pm 0.016$. These values for
$\rho_c$ are much higher than older estimates of $\rho_c$ which were
in the range of 0.41 to 0.57~\cite{old}.

From the estimates of
$\tau_M$ shown in Table~\ref{d3q2}, we see that $\tau_M$ does not
appear to increase with
$L$. It may be that there is no critical slowing for IC dynamics
for the 2-component, $d=3$ WR model as is the case for
the $d=3$ Ising model under IC
dynamics~\cite{MaCh96,ChMaTaCh}.

The fluctuations in the estimator of the fugacity,
$\sigma_z$, decrease with $L$ as $\sigma_z \sim L^{-a}$ with
$a \approx 0.5$ for $d=2$ and $a \approx 0.8$ for $d=3$. The $d=2$
value of $a$ is the same as was found for the $d=2$ Ising model in
the IC ensemble while for $d=3$ it is somewhat larger than the
results obtained for the $d=3$ Ising
model~\cite{ChMaTaCh} where $a = 0.69 \pm 0.01$.  The fact that
$\sigma_z \to 0$ as the system size increases insures that the IC
ensemble is close to the canonical ensemble.

Our results for the 3-component WR
fluid in $d=2$ are summarized in Table~\ref{table:d2q3}.
Using the same finite size scaling analysis we used for the
2-component WR fluid, we find that $D = y_h = 1.842 \pm 0.004$ and
hence
$\beta/\nu \approx 0.16$. This result for $y_h$ is consistent with
our observed value of $\gamma/\nu=1.681 \pm 0.008$. These results
are {\em not} consistent with the corresponding value for the
3-state, $d=2$ Potts model where $y_h=28/15$. Even
more surprising, our estimated value of $y_h$ for the WR fluid is
less than the minimum value of
$y_h$ for any $d=2$ Potts model with a continuous
transition ($y_h \approx 1.86603$ for $q=3.332$). This observation
deserves further study. Is the
3-component, $d=2$ WR fluid outside the Potts
universality classes or are the systematic errors considerably larger
than we have guessed?

We also find that
$\rho$ approaches $\rho_c$ as $1/L$ and extract the
value $\rho_c=1.657 \pm 0.001$. The standard deviation of the
estimator of the fugacity decreases as $\sigma_z \sim L^{-0.4}$.
 A log-log
plot of
$\tau_M$ versus
$L$ yields the estimate of the dynamic exponent $z_M = 0.58$, that
is, for this case $z$ is sufficiently large for us to conclude that
$z > 0$.

Our results for the 3-component, $d=2$ WR fluid are
summarized in Table~\ref{table:d3q3}. The corresponding 3-state,
$d=3$ Potts model is believed to have a first-order
transition, and it is likely that this behavior holds for
the 3-component $d=3$ WR fluid. On the other hand, both the
observed values of $M$ and
$\chi$ are well-described by power laws. This situation also holds
for the 3-state Potts model for computationally accessible
system sizes and reflects the fact that the transition is very weakly
first-order. More study is needed to determine the order of the
transition for this case of the WR model. The IC method finds the
transition temperature for Potts models independently of whether the
transition is first-order or continuous~\cite{MaCh96}. Hence, we
believe that extrapolated values of $z$ yields the transition value
of the fugacity, $z_c \approx 1.16$, and that $\rho \approx 0.795$
lies between the density of the two co-existing phases at the
transition. The standard deviation of the fugacity decreases with
$L$ as $\sigma_z \sim L^{-0.25}$. The effective dynamic exponent
describing the increase in $\tau_M$ is $z_M =0.62$.

We have shown that cluster methods may be effectively used to study
the Widom-Rowlinson model. To our knowledge, the IC algorithm is the
first example of an algorithm that performs efficiently near the
critical point of a continuum system with hard-core interactions. 
We have obtained accurate values of the critical density and
fugacity for Widom-Rowlinson models in $d=2$ and 3. The 2-component
Widom-Rowlinson model appears to be in the Ising universality class.
However, the 3-component $d=2$ model deserves further study and
might not be in the 3-state Potts universality class.

This work was supported by NSF grants DMR-9632898
and DMR-9633385.

\begin{table}
\caption{Dependence of $M$, $\rho$, $\chi$, $z$, $\sigma_z$, and
$\tau_M$ on $L$ for the 2-component, $d=2$ WR fluid. The error
estimates represent one standard deviation. The averages are over
$10^5$ spanning clusters.}
\label{d2q2}
\begin{tabular}{rcccccc}
$L$      & $M$          & $\rho$        & $\chi$        & $z$ & $\sigma_z$
&  $\tau_M$ \\
\tableline
40  &  1511(1)  & 1.5247(4) & 1584(2)  & 1.7201(7) & 0.212 & 0.58
\\
60  &  3233(3)  & 1.5379(3) & 3217(5)  & 1.7247(6) & 0.170 & 0.60
\\
80 &   5527(5)  & 1.5450(3) & 5289(9)  & 1.7267(6) & 0.150 & 0.72
\\
120 & 11836(12)  & 1.5516(3) & 10760(20) & 1.7265(5) & 0.120 & 0.78
\\
160 & 20282(20)  & 1.5552(2) & 17752(30) & 1.7262(4) & 0.101 & 0.77
\\
\end{tabular}
\end{table}

\begin{table}
\caption{Dependence of $M$, $\rho$, $\chi$, $z$, $\sigma_z$, and
$\tau_M$ on $L$ for the 2-component,
$d=3$ WR fluid. The averages are over $10^6$ spanning
clusters.} \label{d3q2}
\begin{tabular}{rcccccc}
$L$     & $M$   & $\rho$        & $\chi$        & $z$ & $\sigma_z$ &
$\tau_M$ \\
\tableline
10 &  313.0(1) & 0.74022(7)    & 119.5(2)  & 0.9387(1) & 0.138 & 0.59
\\
20 & 1745.4(6) & 0.74440(4)  & 466.1(9)  & 0.940(1) & 0.077 & 0.57
\\
30 & 4768(2)   & 0.74567(2)       & 1031(2)   & 0.9403(1) & 0.056 &
0.57
\\
\end{tabular}
\end{table}

\begin{table}
\caption{Dependence of $M$, $\rho$, $\chi$, $z$, $\sigma_z$, and
$\tau_M$ on $L$ for the 3-component, $d=2$ WR fluid.
The averages are over $10^5$ spanning clusters.}
\label{table:d2q3}
\begin{tabular}{rcccccc}
$L$     & $M$   & $\rho$        & $\chi$        & $z$ & $\sigma_z$ &
$\tau_M$ \\
\tableline
40   &  1480(2) & 1.6124(8)     &  1543(4)      & 1.965(3) & 0.364
& 0.88
\\
80        &  5325(7)    & 1.6352(6)     &  4987(13)     & 1.960(1) & 0.280
& 1.2
\\
120      & 11211(18)    & 1.6418(6)     &  9810(30)     & 1.953(1) & 0.237
& 1.7
\\
160      & 19013(32)    & 1.6455(6)     & 15869(50)     & 1.949(1) & 0.213
& 1.9
\\
\end{tabular}
\end{table}

\begin{table}
\caption{Dependence of $M$, $\rho$, $\chi$, $z$, $\sigma_z$, and
$\tau_M$ on $L$ for the 3-component, $d=3$ WR fluid. The
averages are over $10^5$ spanning clusters.}
\label{table:d3q3}
\begin{tabular}{rcccccc}
$L$     & $M$   & $\rho$        & $\chi$        & $z$ & $\sigma_z$ & $\tau_M$\\
\tableline
10      &  299.6(3)     & 0.7914(2)     & 109.8(2)      & 1.1789(8) & 0.230 &
0.58 \\
20      & 1639(2)       & 0.7947(2)     & 409.1(9)      & 1.1717(6) & 0.145 &
0.74 \\
30      & 4407(6)   & 0.7947(1) & 874(2)          & 1.1670(5) & 0.117 & 0.83 \\
\end{tabular}
\end{table}

\begin{figure}[h]
\caption{Log-log plot of $M$, the average mass of the
spanning cluster, versus $L$, the linear dimension of the lattice
for the 2-component, $d=2$ WR fluid. A least squares
fit yields $D=2 - \beta/\nu = 1.873$.}
\label{fig:MLq2d2}
\end{figure}

\begin{figure}[h]
\caption{Plot of $\rho$ versus $1/L$ for the
2-component, $d=3$ WR fluid. A
least squares fit yields $\rho_c=0.7484$.}
\label{fig:rLq2d3}
\end{figure}


\begin{thebibliography}{1}
\bibitem{note}
Alternatively, one of the components can be
integrated out leading to a simple fluid with $N$-body
interactions. The demixing transition
in the 2-component version is a liquid-vapor transition in
the simple fluid.
\end{thebibliography}
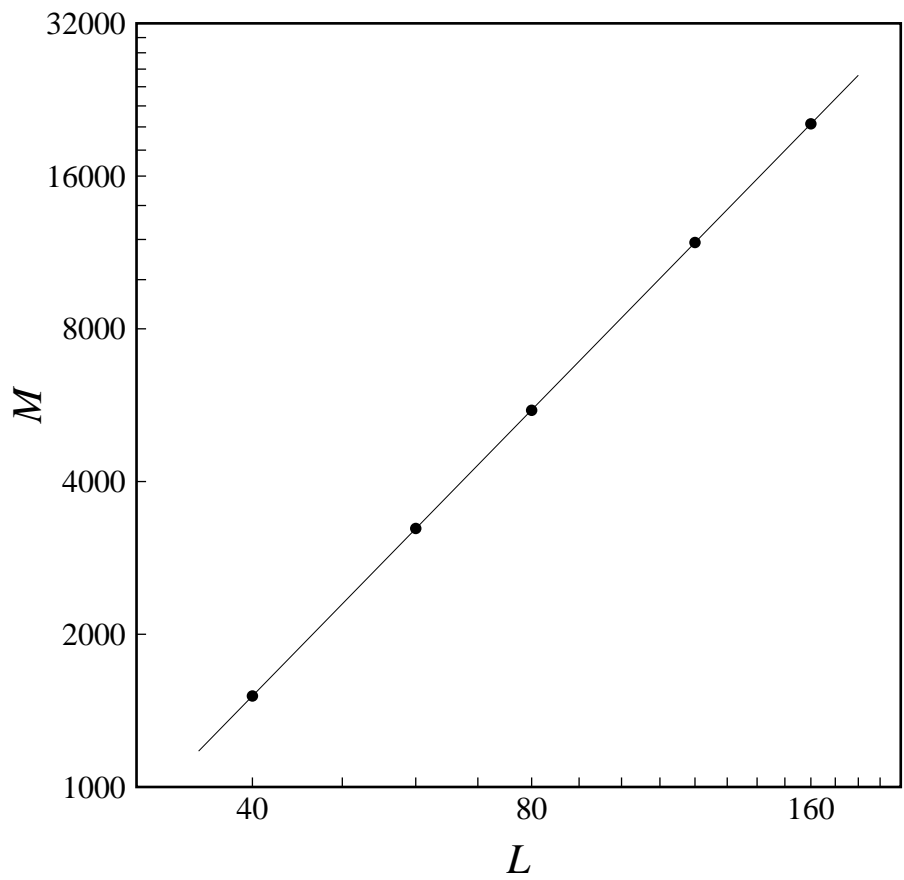
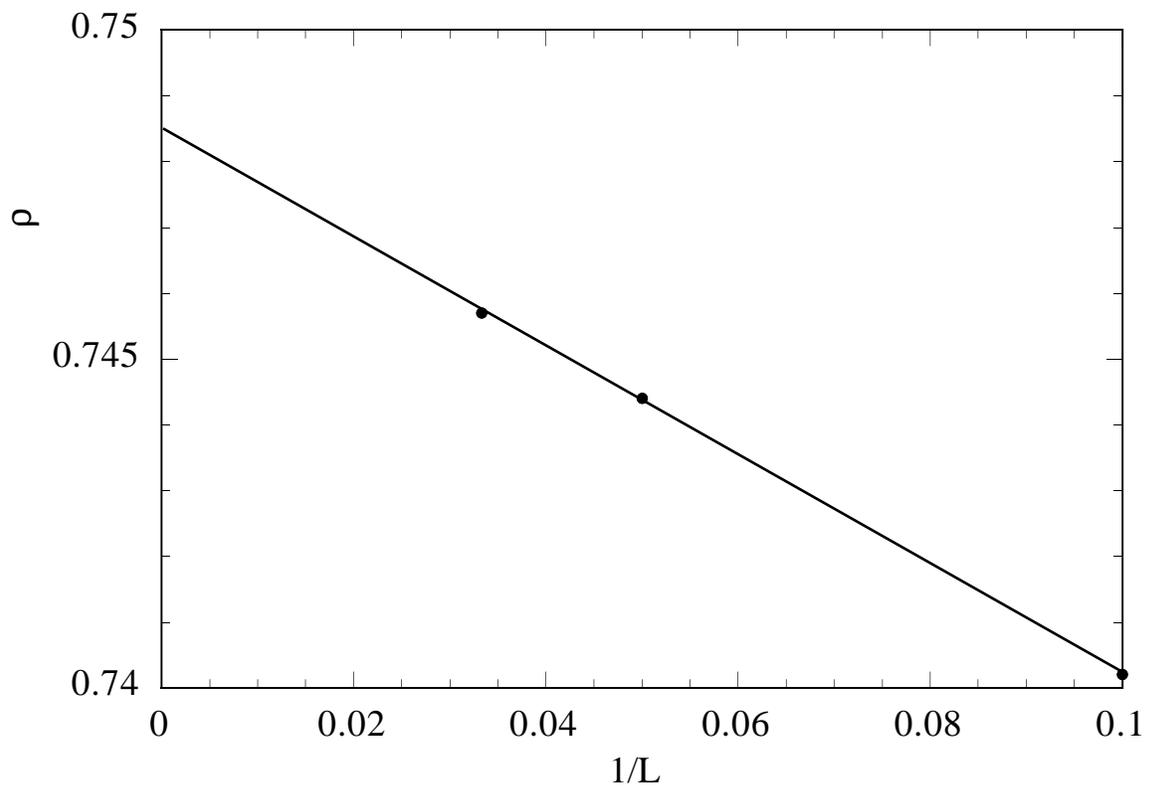
\end{document}